\documentstyle[aps,preprint]{revtex}
\draft
\begin{document}

\title{Superconductive Phonon Anomalies in
High-$T_c$ Cuprates}

\author{B. Normand, H. Kohno and H. Fukuyama}

\address{Department of Physics, University of Tokyo,
7-3-1 Hongo, Bunkyo-ku, Tokyo 113, Japan.}

\date{\today}

\maketitle

\begin{abstract}

	We consider the effects on phonon dynamics of spin-lattice coupling
within the slave-boson mean-field treatment of the extended $t$-$J$ model.
With no additional assumptions the theory is found to give a semi-quantitative
account of the frequency and linewidth anomalies observed by Raman and
neutron scattering for the 340$cm^{-1}$ $B_{1g}$ phonon mode in
$YBa_2Cu_3O_7$ at the superconducting transition. We
discuss the applicability of the model to phonon modes of different
symmetries, and report a connection to spin-gap features observed in
underdoped YBCO. The results suggest the possibility of a unified
understanding of the anomalies in transport, magnetic and lattice properties.

\end{abstract}


\newcommand{\bk}{{\bf k}}
\newcommand{\bx}{{\bf x}}
\newcommand{\bq}{{\bf q}}
\newcommand{\br}{{\bf r}}
\newcommand{\bS}{{\bf S}}
\newcommand{\beq}{\begin{equation}}
\newcommand{\eeq}{\end{equation}}
\newcommand{\bea}{\begin{eqnarray}}
\newcommand{\eea}{\end{eqnarray}}


	Anomalies in the transport and magnetic properties of
high-temperature superconductors, observed in both normal and superconducting
states, have attracted much interest \cite{rrev}, since these phenomena are
considered to be manifestations of a metallic state arising near the Mott
transition due to strong correlations. Although there is as yet no general
consensus on the
appropriate theoretical description of this anomalous metallic state, the
approach based on the $t$-$J$ model \cite{razr} treated by the slave-boson
mean-field theory \cite{rbgks} has given important clues to the
understanding of both transport \cite{rnl} and spin excitation properties
\cite{rTKFs,rTKF,rlw,rul}. These include the temperature-dependences of the
resistivity and Hall coefficient, and the different temperature-dependences
\cite{ry,rr} of the shift and rate of nuclear magnetic resonance between
high- and low-doping regions, especially the ``spin-gap'' phenomenon
\cite{rTKF,rul} first noted by Yasuoka \cite{ry}. There have
been various experimental reports
\cite{rliv,ralt,rham} which indicate anomalous temperature-dependences
also in the frequency shift and linewidth of phonons around the
superconducting transition. Further, studies by neutron scattering
\cite{rtejs,rarai}, EXAFS \cite{roya}, ion-channeling \cite{ryam} and
ultrasound measurements \cite{rnsmftk}
all provide strong evidence of a link between lattice
anomalies and superconductivity. Although some aspects of the phonon
problem have been studied theoretically \cite{rzz}, to our knowledge
there have not so far been any efforts to understand all of these
interesting low-lying excitations on a unified basis.

	We consider the spin-phonon coupling arising naturally within the
extended $t$-$J$ model of a single $CuO_2$ layer, which has been shown
\cite{rTKFs,rTKF} at the
mean-field level to give a good account of many features of the spin
excitations in both $La_{2-x}Sr_xCuO_4$ (LSCO) and $YBa_2Cu_3O_{7-\delta}$
(YBCO). The Hamiltonian is
\beq
H = - \sum_{ij} t_{ij} a_{i s}^{\dag} a_{j s} + \sum_{ \langle ij \rangle }
J_{ij} \bS_i {\bf {.S}}_j ,
\label{eh}
\eeq
where the Hilbert space is that without double occupancy, $t_{ij}$ corresponds
to the transfer integrals used to reproduce the Fermi surface, and the
superexchange interaction $J_{ij}$, which is assumed to be finite only between
nearest neighbours, has been taken as a constant in previous treatments.

	In YBCO, the $Cu O_2$ layer is ``buckled'', by which is meant
that the oxygen atoms $O(2)$ and $O(3)$ lie out of the plane of the $Cu$
atoms, as shown schematically in Fig.~1. Then $t_{ij}$ and $J_{ij}$
between the nearest-neighbor $Cu$ sites have contributions linear in the
magnitude of the oxygen displacement along the $c$-axis, $u_{i}^{\alpha}$
($i$ refers to the $Cu$ sites on the square lattice, and $\alpha$ to either
$O(2)$ ($\alpha = x$) or $O(3)$ ($\alpha = y$)), given by $t_{ij} = t [ 1
- \lambda_t (u_{i}^{\alpha}/a) ]$, where $a$ and $\lambda_t$ are the
distance between $Cu$ sites and the coupling constant, respectively, and
similarly for $J_{ij}$ with coupling $\lambda_J$. A microscopic
estimate of $\lambda_t$ and $\lambda_J$ requires several steps.
{}From the structural data of Ref. \cite{rgui}, the equilibrium $O$
displacement
is $u_0 = 0.256 {\rm {\AA}}$ (neglecting henceforth the 5\% anisotropy
between $\hat{x}$ and $\hat{y}$); although the degree of buckling is small,
$u_0/a \ll 1$, its inclusion is crucial in providing a coupling which is
strong and linear.  Following Ref. \cite{rHarrison},
the dependence on interatomic separation of the transfer integral
$t_{\sigma}$ ($t_{\pi}$) between $Cu$:$d_{x^2 - y^2}$ and $O$:$p_{\sigma}$
($O$:$p_{\pi}$) is given by
\bea
t_{\sigma} & = & \frac{\sqrt{3}}{2} V_{pd \sigma} \left[ 1 -
\frac{3}{2} \left( \frac{2 u}{a} \right)^2 \right] + \left( \frac{2 u}{a}
\right)^2 V_{pd \pi} \label{etsp} \\
t_{\pi} & = & \frac{\sqrt{3}}{2} \frac{2 u}{a} V_{pd \sigma} \left(
1 - \frac{2}{\sqrt{3}} \frac{V_{pd \pi}}{V_{pd \sigma}} \right) ,
\eea
where $\mbox{$\frac{\sqrt{3}}{2}$} V_{pd \sigma}$ ($V_{pd \pi}$) is the
transfer integral between $d_{x^2 - y^2}$ ($d_{xz}$) and $p_{\sigma}$
($p_{z}$) orbitals with separation $d = \left( (\mbox{$\frac{1}{2}$} a)^2 +
u^2 \right)^{1/2}$ along the ${\hat x}$-direction. Taking $V_{pd} (d)
\propto d^{- 7/2}$ and writing $u = u_0 + \delta u$, where $\delta u$
is the oscillation amplitude, gives $t_{\sigma} (u) / t_{\sigma} (u_0)
= 1 - 2.03 \delta u / a$ and $t_{\pi} (u) / t_{\sigma} (u_0) = 1.53 \delta
u / a$. The final requirement is the dependence
on $t_{\sigma}$ and $t_{\pi}$ of $t_{ij}$ and $J_{ij}$.
For $J_{ij}$, the perturbative expression $J = 4 \left( t_{\sigma}^4
- 2 t_{\sigma}^2 t_{\pi}^2 \right) / {\overline {\Delta}}^2 {\overline {U}}$
({\sl {cf.}} \cite{rej}) would give a large coupling constant $\lambda =
10.4$. However, use of the lowest-order perturbation form
is not well justified, and investigations of the influence of
higher-order terms find for the parameters of the $Cu O_2$ layer the
effective relationships $J \propto t_{pd}^x$ with $x \simeq 2.3$ \cite{rej},
and $t \propto t_{pd}^y$ with $y \simeq 1.0$ \cite{repc}; the result for $y$
requires consideration of the many inter-site transfer integrals which
contribute to the hopping of a Zhang-Rice singlet. The exact values of
$x$ and $y$, and particularly of the powers $y^{\prime}$ and $y^{\prime
\prime}$ corresponding to the extended hopping terms
$t^{\prime}$ and $t^{\prime \prime}$, are quite sensitive to both the
parameter choice and to the symmetry of the phonon mode \cite{repc}.
This treatment disregards changes in ${\overline {\Delta}}$, which we
believe to be appropriate for the local, screened deformation processes
under investigation. By contrast, a $d$-dependence of ${\overline {\Delta}}$
is required to account for the weak relation $J \propto d^{- \alpha}$, $4 <
\alpha < 6$, found by pressure- and substitution-induced variation of the
bond length \cite{rbld},
where in addition the Madelung energy is altered \cite{rsa}.
Mindful of these uncertainties, we proceed by taking the values $x = 2.0$,
$y = 1.0$ so that $\lambda_J = 5.2$ and $\lambda_t = 2.6 = \mbox{$\frac{1}
{2}$} \lambda_J$, and neglect
modulation of the extended $t_{ij}$ terms for the purposes of the current
study. These estimates have also been found to give good agreement with
the measured isotope shift in YBCO \cite{rzea}, a result (deferred to a
future publication) which provides an independent indication of their
validity.

	The terms in $H$ (Eq. (\ref{eh})) describing the coupling of
$u_{i}^{\alpha}$ to the spin degrees of freedom may be rewritten in the
slave-boson, mean-field approximation as
\bea
& & - \sum_{i} \sum_{\alpha = x,y} \left\{ t \lambda_t \mbox{$\left(
\frac{u_i^{\alpha}}{a} \right)$} \langle b_i b_{i + \alpha}^{\dag} \rangle
\chi_{i,i + \alpha} \right. \label{ejss} \\ & &
\left. + \mbox{$\frac{3}{8}$} J \lambda_J \mbox{$\left(
\frac{u_i^{\alpha}}{a} \right)$} \left[ \langle \chi_{i,i + \alpha}^{\dag}
\rangle \chi_{i,i + \alpha} + 2 \langle \Delta_{ij}^{\dag} \rangle
\Delta_{ij} + h.c. \right] \right\} \nonumber ,
\eea
where $\chi_{i,j} = \sum_s f_{i s}^{\dag} f_{j s}$ and $\Delta_{ij} = \left(
f_{i \uparrow} f_{j \downarrow} - f_{i \downarrow} f_{j \uparrow} \right)
/ \sqrt{2} $.
We do not consider the phonon-holon coupling vertex which is also
contained in the $t$ term, because in the normal state this will vanish
at $q = 0$, while in the Bose-condensed state the holon has no dynamics.

	Here we will be concerned mainly with the $T$-dependence
of the frequency shift and linewidth of the $340cm^{-1}$ $B_{1g}$ mode
in $YBa_2Cu_3O_7$, which is an out-of-phase oscillatory motion of only
the planar oxygen atoms ($u_{i}^x = - u_{i}^y$), and is also out of phase
between the planes of the bilayer \cite{rtc}. This mode has attracted
experimental interest because it shows the largest effects at the
superconducting transition, and there are available detailed Raman
\cite{rliv,ralt,rham} ($q = 0$) and inelastic neutron scattering
\cite{rprperh} (also $q \ne 0$) data. We will also show results for
$A_{1g}$- or $A_{2u}$-symmetric (in-phase, $u_{i}^x = u_{i}^y$)
oscillations, and discuss the relevance of the model for these.

	The effect of the coupling on the dynamical properties of the
phonon is calculated from the lowest-order spinon correction to the phonon
self-energy $\Pi_{ph}$ (Fig.~2). Second-order perturbation theory in terms of
the coupling results in a frequency shift $\delta \omega = ${\sl Re}$\,
\Pi_{ph}$ for $|\delta \omega| \ll \omega_0$, which is given at $q = 0$ by
\beq
\delta \omega = c \left( \lambda_J J \right)^2 \frac{4}{N} \sum_{k} B_k
\frac{1}{\omega^2 - \left( 2 E_k \right)^2} \frac{\tanh \left( \frac{E_k}{2 T}
\right)}{E_k} ,
\label{edo}
\eeq
where $c = \mbox{$\left( \frac{3}{4 a} \right)^2$} \langle u^2 \rangle =
1.18 \times 10^{-4}$, $\langle u^2 \rangle = \mbox{$\frac{\hbar}{2 M
\omega_0}$} = ( 0.055 {\rm {\AA}} )^2$, in which $\omega_0 = 340 cm^{-1}$
is taken for the phonon frequency of interest and $M$ is the mass of the
$O$ atom, $E_k$ is defined by $E_k = \left[ \xi_{k}^2 + \Delta_{k}^2
\right]^{1/2}$, with $\xi_k$ the spinon band energy relative to the
chemical potential and $\Delta_k = - \mbox{$\frac{3 \sqrt{2}}{4}$}
J \Delta (\cos k_x - \cos k_y)$ the singlet order
parameter, and $B_k$ is the form factor
\bea
B_k & = & 2 \Delta^2 \left( \gamma_k \xi_{k} + \mbox{$\frac{3 J}{4}$}
{\bar {\chi}} \eta_{k}^2 \right)^2 \;\;\;\;\;\;\;\;\;\; B_{1g}, \, B_{2u}
\; {\rm {modes}} \label{eff} \nonumber \\ B_k & = &
2 \Delta^2 \eta_{k}^2 \left( \xi_{k} + \mbox{$\frac{3 J}{4}$}
{\bar {\chi}} \gamma_k \right)^2
\;\;\;\;\;\;\;\;\;\; A_{1g}, \, A_{2u} \; {\rm {modes}} ,
\eea
with $\gamma_k = \cos k_x + \cos k_y$, $\eta_k = \cos k_x - \cos k_y$,
$\Delta = \langle \Delta_{ij} \rangle$ and ${\bar {\chi}} \equiv
\langle \chi_{ij} \rangle + \frac{2 t \delta}{3 J}$.
One observes in Eq. (\ref{eff}) that $B_k \propto \Delta^2$, and so there
is no phonon energy correction due to spin coupling in the normal state; this
result is a special feature of the $\omega \ne 0$ phonon mode we
consider at $q = 0$, and of the current level of approximation.
The superconductivity-induced correction to the linewidth $\Gamma$ is
computed similarly from {\sl Im}$ \, \Pi_{ph}$.

	A detailed numerical analysis of the self-consistent mean-field
equations has been carried out \cite{rTKFs} to compute the
temperature-dependences of $\Delta$, ${\bar {\chi}}$ and chemical potential
for several choices of doping $\delta$, for the transfer integrals
$t = 4 J$, $t^{\prime} = - \mbox{$\frac{1}{6}$} t$ and
$t^{\prime \prime} = \mbox{$\frac{1}{5}$} t$, appropriate to YBCO.
With these parameters, the frequency-dependence of $\delta \omega$, given by
Eq. (\ref{edo}), has been evaluated numerically for several temperatures,
and the results are shown in Fig.~3 for $q = 0$ and $\delta = 0.2$ at
$T/T_{RVB} = 0.2$, where $T_{RVB} = 0.069 J$ is the onset temperature for
the singlet RVB order parameter $\Delta$.
All calculations were performed with an assumed Lorentzian broadening
of the spinon spectrum $\Gamma = 0.12 k_B T_{RVB}$, and are found to be
quite insensitive to the value of this parameter.
{}From the $\omega$-dependent denominator in Eq. (\ref{edo}), the frequency
where $\delta \omega$ of the $B_{1g}$ mode changes sign is an approximate
measure of the value of $2 \Delta_k (T)$ near the $(\pi,0)$ points,
where the gap is maximal: at higher temperatures
(but below $T_{RVB}$) the $\omega$-dependence of $\delta \omega$
is qualitatively the same, but the sign-change occurs at lower
frequencies. The experimental mode frequency $\omega_0$ for the
$B_{1g}$ mode is near, but just below, the low-$T$ crossover, so this mode can
be expected to show a maximal effect. The linewidth broadening $\delta
\Gamma$ has the form to be expected for the corresponding imaginary part,
namely a peak at the crossover frequency.

	The temperature-dependence of $\delta \omega$ is shown in
Fig.~4(a) for $q = 0$ and $\omega / J = 0.25$, a value
in reasonable quantitative agreement with the
340$cm^{-1}$ mode: while modes with frequencies considerably less than
$2 \Delta$ have a sharp transition, those close to it show clearly
that as $\omega$ approaches the maximal value of $2 \Delta_k
(T = 0)$, the shift in frequency occurs at a temperature
somewhat below $T_{RVB}$.
This corresponds well to the observations of Ref. \cite{ralt}, where the
full frequency shift develops over a range of temperatures below the onset.
Comparison with the experimental result \cite{ralt} for the broadening of the
$B_{1g}$ mode again shows a good agreement in sign and magnitude ($\delta
\omega \simeq \delta \Gamma \simeq 0.01 \omega_0$) for a frequency close to
$\omega_0$ (Fig.~4(b)), while at lower frequencies (``off resonance'')
$\delta \Gamma$ is suppressed.
We have extended our analysis to the case where the phonon has a
finite wavevector $\bq$, finding smaller but sharper anomalies in further
qualitative agreement with experiment; these calculations will be presented
elsewhere. Quantitatively, the magnitude of the effects given by the model
with the chosen values of $\lambda_t$ and $\lambda_J$ is within a factor
of 1.5 of the Raman measurements \cite{ralt} on $Y Ba_2 Cu_3 O_7$;
this degree of correspondence, as well as that
in $\omega_0$ above, can be regarded as satisfactory within a mean-field
treatment using no adjustable parameters.

	In Figs.~3 and 4 are shown not just the $B_{1g}$ mode, but also the
results for a mode of in-phase $O(2)$ and $O(3)$ oscillations, which would
correspond to $A_{1g}$ (Raman active, $440cm^{-1}$) or $A_{2u}$
(infrared-active, $307cm^{-1}$) symmetry. At the level of
the current approximation these have negligible anomalies, a qualitative
difference from $B_{1g}$ which arises from the form factors $B_k$ (\ref{eff})
with $d$-wave singlet pairing; we note that extended $s$-pairing gives an
immeasurably small frequency shift for both types of mode symmetry. However,
experimental studies of the $A_{2u}$ mode \cite{rliv},
whose frequency is close to that investigated, reveal a strong shift $\delta
\omega / \omega_0 \simeq 1 \%$. We believe that the model
cannot reproduce this result because
it contains nothing to account for the charge transfer between the $Cu O_2$
planes of a bilayer which accompany this mode in YBCO,
and so is inapplicable. Developing the theory to include interplane
charge motion \cite{rhfk} is beyond the basic $t$-$J$
framework, and will be pursued in a subsequent publication.

	The 193$cm^{-1}$ $B_{2u}$ mode is an out-of-phase oscillation
of planar $O$ (in-phase between the planes of the
bilayer) involving little charge motion, and thus a single-layer model
is expected to be valid. Because this mode is i.r.-silent,
it may be studied only by neutron scattering, and investigations on
$Y Ba_2 Cu_3 O_7$ have recently been performed \cite{rhs}.
The observed anomalies are sharp and occur close to $T_c$,
with a relative magnitude $\delta \omega / \omega_0 \simeq 1\%$ similar
to those of the $B_{1g}$ mode, features which are indeed well described
by the current model at this lower frequency.
	Finally, the effects studied should be strongly suppressed in the
$E$-symmetric phonon modes, where $O$ displacements are parallel
to the plane and so have little influence on the interaction $J_{ij}$, a
feature also largely in agreement with experiment.

	These results
raise the interesting possibility of probing the spin-gap behavior
observed in the temperature-dependence of the NMR relaxation rate by
considering the frequency shifts of particular phonons in members of the
YBCO class in the low-doping regime, where the rate has been found to be
maximal at some value $T_0$ above the superconducting critical temperature
$T_c$. There exist already several experimental reports \cite{rliv,ralt,rham}
of anomalies in the frequency shift well above $T_c$, and near the temperature
where the NMR rate exhibits a maximum, while recent, highly accurate Raman
studies of phonon anomalies in underdoped YBCO compounds \cite{rklbjb}
show clearly the onset of a frequency shift
at some $T_0 \simeq 150K$, followed by growth of this
shift as temperature is lowered, until a saturation below $T_c$.
Such features may be understood on the basis of the mean-field phase diagram
of the extended $t$-$J$ model \cite{rTKF}, in which $T_{RVB}$ is indeed higher
than $T_c$ only in the low-doping region, and they would be interpreted
as the onset of singlet-RVB order around $T_0$, {\it i.e.} its identification
with the crossover temperature $T_{RVB}$ \cite{rnl}.
This result is consistent with the observation that
the specific heat anomalies at $T_c$ in these cases are qualitatively
different from those near optimal doping \cite{rlsh,rskhs}.
We suggest that the existence of two temperature scales
is the most likely explanation for the contrasting low-$q$ behavior
of the energy shifts in nominal $O_{6.92}$ and $O_7$ compounds observed in
Ref. \cite{rprperh}.

	In summary, we have proposed a theory of spin-phonon coupling
which accounts very well for the anomalies observed in
phonon modes in typical high-$T_c$ cuprates
of the YBCO class, based on the mean-field approximation to the
extended $t$-$J$ model. We believe that this model is
a useful step in constructing a coherent theory of the spin, transport
and lattice properties in the anomalous metallic state,
which has a strong bearing on high-temperature superconductivity.

	We are grateful to M. Arai, M. Cardona, Y. Endoh, H. Eskes,
H. Kino, N. Nagaosa, H. Oyanagi, Y. Tokura, H. Yasuoka and especially
to B. Batlogg and M. Sato for helpful discussions.
This work was supported financially by the Monbusho International
Scientific Research Program No. 05044037 and the Grant-in-Aid for
Scientific Research No. 04240103 of the Ministry of Education, Science and
Culture, Japan. B.N. wishes to acknowledge the support of the Japan
Society for the Promotion of Science.

\section*{Figure Captions}

\bigskip
\noindent
Fig. 1: Schematic representation of $CuO_2$ layer, showing
``buckling'' deformation of the equilibrium positions of
$O(2)$ and $O(3)$ atoms out of the plane of the $Cu$ atoms, appropriate
for YBCO. $a$, $b$ and $c$ represent crystal axes.

\bigskip
\noindent
Fig. 2: Diagrammatic representation of the lowest-order contribution
to the phonon self-energy $\Pi_{ph}$ due to coupling to spinons.

\bigskip
\noindent
Fig. 3: Phonon frequency shift $\delta \omega$ for $B_{1g}$ ($\circ$)
and $A_{2u}$ ($\times$) modes at $q = 0$, as a function of frequency at
$T = 0.2 T_{RVB}$. The arrow indicates the frequency whose
temperature-dependence is illustrated in Fig.~4.

\bigskip
\noindent
Fig. 4: Phonon frequency shift $\delta \omega$ (a) and linewidth
broadening $\delta \Gamma$ (b) for $B_{1g}$ ($\circ$) and $A_{2u}$
($\times$) modes at $q = 0$, as functions of temperature for mode
frequency $\omega_0 = 0.25 J$.


\begin{thebibliography}{99}

\bibitem{rrev} See for example {\sl {J. Low Temp. Phys.}} {\bf {95}},
Nos. 1/2 (1994)

\bibitem{razr} P. W. Anderson, {\sl {Science}} {\bf {235}}, 1196 (1987);
F.-C. Zhang and T. M. Rice, {\sl {Phys. Rev.}} {\bf {B 37}}, 3759 (1988)

\bibitem{rbgks} G. Baskaran, Z. Zou and P. W. Anderson, {\sl {Solid State
Commun.}} {\bf {63}}, 973 (1987); C. Gros, R. Joynt and T. M. Rice,
{\sl {Phys. Rev.}} {\bf {B 36}}, 8190 (1987); G. Kotliar and J. Liu,
{\sl {Phys. Rev.}} {\bf {B 38}}, 5142 (1988); Y. Suzumura, Y. Hasegawa
and H. Fukuyama, {\sl {J. Phys. Soc. Jpn.}} {\bf {57}}, 2768 (1988)

\bibitem{rnl} P. A. Lee and N. Nagaosa, {\sl {Phys. Rev.}}
{\bf {B 46}}, 5621 (1992)

\bibitem{rTKFs} T. Tanamoto, H. Kohno and H. Fukuyama, {\sl {J. Phys. Soc.
Jpn.}} {\bf {62}}, 717 (1993)

\bibitem{rTKF} T. Tanamoto, H. Kohno and H. Fukuyama, {\sl {J. Phys. Soc.
Jpn.}} {\bf {62}}, 1455 (1993) and {\bf {63}}, 2741 (1994)

\bibitem{rlw} M. Lercher and J. Wheatley, {\sl {Phys. Rev.}}
{\bf {B 49}}, 736 (1994)

\bibitem{rul} M. Ubbens and P. A. Lee, {\sl {Phys. Rev.}}
{\bf {B 50}}, 438 (1994)

\bibitem{ry} H. Yasuoka, T. Imai and T. Shimizu, in {\it Strong
Correlations and Superconductivity}, eds. H. Fukuyama, S. Maekawa and A. P.
Malozemoff, Springer Verlag, 254 (1989).

\bibitem{rr} T. M. Rice, in {\sl {The Physics and Chemistry of Oxide
Superconductors}}, ed. Y. Iye and H. Yasuoka (Springer Verlag, Berlin, 1992)

\bibitem{rliv} A. P. Litvinchuk, C. Thomsen and M. Cardona, {\sl {Solid
State Commun.}} {\bf {83}}, 343 (1993)

\bibitem{ralt} E. Altendorf, X. K. Chen, J. C. Irwin, R. Liang and
W. N. Hardy, {\sl {Phys. Rev.}} {\bf {B 47}}, 8140 (1993)

\bibitem{rham} K.-M. Ham, J.-T. Kim, R. Sooryakumar and T. R. Lemberger,
{\sl {Phys. Rev.}} {\bf {B 47}}, 11439 (1993)

\bibitem{rtejs} B. H. Toby, T. Egami, J. D. Jorgensen and M. A.
Subramanian, {\sl {Phys. Rev. Lett.}} {\bf {64}}, 2414 (1990)

\bibitem{rarai} M. Arai, K. Yamada, Y. Hidaka, S. Itoh, Z. A. Bowden,
A. D. Taylor and Y. Endoh, {\sl {Phys. Rev. Lett.}} {\bf {69}}, 359 (1992)

\bibitem{roya} S. D. Conradson and I. D. Raistrick, {\sl {Science}}
{\bf {243}}, 1340 (1989)

\bibitem{ryam} T. Haga, K. Yamaya, Y. Abe, Y. Tajima and Y. Hidaka,
{\sl {Phys. Rev.}} {\bf {B 41}}, 826 (1990)

\bibitem{rnsmftk} M. Nohara, T. Suzuki, Y. Maeno, T. Fujita, I. Tanaka
and H. Kojima, {\sl {Phys. Rev. Lett.}} {\bf {70}}, 3447 (1993)

\bibitem{rzz} R. Zeyher and G. Zwicknagel, {\sl {Z. Phys.}} {\bf {B 78}},
175 (1990)

\bibitem{rgui} M. Guillaume, P. Allenspach, J. Mesot, B. Roessli, U. Staub,
P. Fischer and A. Furrer, {\sl {Z. Phys.}} {\bf {B 90}}, 13 (1993)

\bibitem{rHarrison} W. A. Harrison, {\sl {Electronic Structure and the
Properties of Solids}} (W. H. Freeman and Company, San Francisco, 1980)

\bibitem{rej} H. Eskes and J. H. Jefferson, {\sl {Phys. Rev.}}
{\bf {B 48}}, 9788 (1993)

\bibitem{repc} H. Eskes, private communication

\bibitem{rbld} V. V. Struzhkin, U. Schwarz, H. Wilhelm and K. Syassen,
{\sl {Mat. Sci and Eng.}} {\bf {A 168}}, 103 (1993); Y. Ohta, T. Tohyama
and S. Maekawa, {\sl {Phys. Rev. Lett.}} {\bf {66}}, 1226 (1991)

\bibitem{rsa} We are grateful to G. Sawatzky and T. Arima for advice on
this point.

\bibitem{rzea} D. Zech, H. Keller, K. A. M\"uller, K. Conder, E. Kaldis,
E. Liarokapis and N. Poulakis, to appear in {\sl {Nature}} (1994)

\bibitem{rtc} For a review of phonon modes and nomenclature see
C. Thomsen and M. Cardona, in {\sl {Physical Properties
of High Temperature Superconductors I}}, ed. D. Ginsberg (World Scientific,
Singapore, 1989)

\bibitem{rprperh} N. Pyka, W. Reichardt, L. Pintschovius, G. Engel,
J. Rossat-Mignod and J. Y. Henry, {\sl {Phys. Rev. Lett. }}
{\bf {70}}, 1457 (1993)

\bibitem{rhfk} G. Hastreiter, F. Forsthofer and J. Keller, {\sl {Solid
State Commun.}} {\bf {88}}, 769 (1993)

\bibitem{rhs} H. Harashina {\sl {et al.}}, unpublished

\bibitem{rklbjb} M. K\"all, A. P. Litvinchuk, P. Berastegui, L.-G.
Johansson and L. B\"orjesson, {\sl {Physica}} {\bf {C 225}}, 317 (1994)

\bibitem{rlsh} J. W. Loram, K. A. Mirza, J. R. Cooper and W. Y. Liang,
{\sl {Phys. Rev. Lett.}} {\bf {71}}, 1740 (1993)

\bibitem{rskhs} S. Shamoto, T. Kiyokura, H. Harashina and M. Sato,
{\sl {J. Phys. Soc. Jpn.}} {\bf {63}}, 2226 (1994)

\end{thebibliography}
\end{document}